\documentclass[12pt, preprint,showkeys, nofootinbib, superscriptaddress]{revtex4-2}
\usepackage{graphics,epsfig}
\usepackage{epstopdf}
\usepackage{graphicx}
\usepackage{dcolumn}
\usepackage{multirow}
\usepackage{amsmath}
\usepackage{diagbox}
\usepackage{footnote}
\usepackage{enumitem}
\begin{document}

\title{Strong Gravitational Lensing by Bardeen Black Hole in Cloud of Strings}

\author{Bijendra Kumar Vishvakarma}
\email{bkv1043@gmail.com}
\affiliation{Department of Physics, Institute of Science, Banaras Hindu University, Varanasi-221005, India}

\author{Shubham Kala}
\email{shubhamkala871@gmail.com}
\affiliation{The Institute of Mathematical Sciences 
\\C.I.T. Campus, Taramani-600113, India}

\author{Sanjay Siwach}
\email{sksiwach@hotmail.com}
\affiliation{Department of Physics, Institute of Science, Banaras Hindu University, Varanasi-221005, India}

\begin{abstract}
We investigate the gravitational lensing by Bardeen black hole in cloud of strings (CoS) in strong field limit. The effect of CoS parameter $b$ has been outlined in comparison with Bardeen black hole lens. The strong deflection limit coefficients are determined in terms of impact parameter for various values of CoS parameter. We obtain magnification of relativistic images and determine relativistic Einstein rings by using the parameters of two astrophysical black hole lenses $SgrA^{*}$ and $M87^{*}$. We constrain CoS parameter of Bardeen black hole using EHT observations for these black holes.
\vspace{2.0cm}
\end{abstract}


\maketitle

\newpage

\section{Introduction}
\label{sec1}
  
 Black holes are known to be singular solutions of the General theory of relativity (GTR). The well known examples are Schwarzschild and Kerr black holes \cite{SW,RP,KERR:1963,book,JMM}. 
 Several charged solutions of gravity coupled with electrodynamics are also known. Investigation of particle
motion (including photon) around black hole has been used widely for indirect evidence of the existence of astrophysical black holes. Black hole imaging using Event Horizon Telescope (EHT) has provided a direct method to see the black holes in the extreme gravity beyond doubt \cite{EHT:2019a,EHT:2022a,EHT:2022f,EHT:2024}. EHT and LIGO-Virgo-KAGRA  results are also used to constrain rotating and charged black hole metrics \cite{Vagnozzi:2023,ABH,PBH,Yuan:2021,Dimitrios:2020}. 

 Bardeen black hole is a solution of Einstein's gravity coupled with non-linear electrodynamics and regarded as a candidate of non-singular spacetime \cite{Bardeen,Dymnikova}. It is a magnetically charged solution of a self gravitating magnetic mono-pole and provides an opportunity to explore the fate of magnetic monopoles in the early universe \cite{Ayon,AyonBeato}. It appears that the fundamental ingredients of nature are one-dimensional strings rather than point-like particles. The early universe may have been dominated by fundamental strings or we can also think of primordial astrophysical objects surrounded by cosmic strings \cite{Bronnikov,Rodrigues:2018,Primordial,CS:Primordial,Alexander:2021}. The primordial black hole surrounded by cloud of strings (CoS) may be an interesting subject to explore the possible signatures of string theory in early universe \cite{Letelier:1979,Damour:2000,Estanislao:2010,Toledoa:2018,Rodrigues:2022}. The inclusion of a string cloud in our model provides a general framework to study deviations from standard black hole physics, which could be due to exotic matter or early-universe relics. Although many string cloud configurations may have dissipated or been restructured during cosmic evolution, there are theoretical models and scenarios in which remnants of such clouds could survive \cite{Patricio:1983,Hindmarsh:1994,Davis:2005,Lorenz:2010,Ganguly:2014cqa,Vachaspati:2015,Pierre:2020}. In this context, we studied several aspects of Bardeen black hole in cloud of strings recently\cite{BKV:2023}.

The phenomenon of gravitational lensing (GL) is used to test the predictions of extreme gravity as well as nature of cosmic objects \cite{Positivist:2024,Refsdal,GL:1992,SV:2007,BHGL,A.Einstein, Liebes:1964}. The weak field limit of bending angle and time delay of spherically symmetric charged black hole in four dimension has been studied using perturbative method of gravitational lensing \cite{Saha:2024}. The strong deflection
of radiation from star is studied by \cite{Jia:2020} and for a black hole by \cite{Chitre:1998,KS:2000,KS:2009,Bozza:2005,GS:2008,Bozza:2002,Bozza:2008,He:2024}.  The analysis is not limited to gravity solution of GTR but has also been used widely for theories beyond GTR \cite{Bozza:2003}. Weak gravitational lensing of Schwarzschild black hole has been investigated in cloud of strings by G. Mustafa et. al., other \cite{Mustafa,Soares:2003} and for various generalizations by \cite{Tsukamoto,Allahyari:2020,Kala:2020,Ovgun:2019,FSchmit,Bhadra,Jha:2022,Molla}. Also weak gravitational lensing of regular and Bardeen black hole in cloud of strings have been studied by \cite{Narzilloev,F:2023}. The lensing in the strong deflection limit (SDL) for a general spherically symmetric spacetime is analyzed by \cite{KS:2000}. The analysis is extended to more general black hole in
GTR and in theories beyond GTR \cite{Bozza:2010,Bozza:2004,Hossein:2016, NUMolla:2024,Ramadhan,AKumar:2003,Xiao-Mei:2022,Chen:2022, Kalita:2023,Ednaldo, Ernesto,CS:2004,Paul:2020,RKumar:2020}. Here, we consider regular Bardeen black hole in cloud of strings and investigate its strong gravitational lensing.
 
 The paper is structured as follows. In section (2) we introduce Bardeen black hole in cloud of strings, its event horizon, and limit on parameter space for the existence of the black hole horizon. Geodesic motion and impact parameter is also introduced to analyze the lensing phenomenon in this section. We also review the strong deflection limit of Bardeen black hole in CoS as lens. In section 3, we determine the magnification and relativistic Einstein ring (first introduced by Virbhadra and Ellis \cite{KS:2000}) in CoS and also compare our results with the parameter space of the well known black holes $M87^{*}$ and $SgrA^{*}$ and constrain the cloud of strings parameter. In section 4 we summarize our results and outline the futuristic directions for further work.  
 
\section{Strong Gravitational lensing from Bardeen Black Hole}
Let us consider the Latelier-Bardeen black hole  which is a solution of Einstein gravity coupled with nonlinear electrodynamics and cloud of strings as a source. Latelier introduced cloud of strings to mimic the environment in early universe \cite{Letelier:1979}. Magnetically charged Bardeen black hole in cloud of strings is known as Latelier-Bardeen black hole \cite{Rodrigues:2022}. The line element of the general spherically symmetric spacetime can be written in the following form,
\begin{equation}
ds^2=-A(r)dt^2+ B(r)dr^2+ C(r) (d\theta^2+\sin^2\theta d\phi^2),
\label{solution}
\end{equation} 
\begin{align*}
    A(r)&=B(r)^{-1}=1-b-\frac{2Mr^2}{(r^2+g^2)^\frac{3}{2}} \hspace{0.2cm} \& \hspace{0.2cm}  C(r)=r^2.
    \end{align*}
     Here, $M$ is the mass of the black hole and the parameters $g$ and $b$ are identified with magnetic monopole charge and CoS parameter respectively. We use natural unit $(G=1,\hbar=1,\& ~ c=1)$ in this paper.
     
  \text{The Event horizon of the 
  black hole metric is given by the roots of $A(r)=0$} 
  and is located at, 
  
\begin{align*}
     r_{h}&=\frac{1}{3\sqrt{3}(1-b)}\sqrt{M^2(9M^2-8)(2^{11}\mathcal{B}^{-1})^\frac{1}{3}+9(2\mathcal{B})^\frac{1}{3}+36M^2-16},
\end{align*}
 where
\begin{align*}
          \mathcal{B}&=3\sqrt{81(1-b)^8g^8M^4-48(1-b)^6g^6M^6}+27(1-b)^4g^4M^4+32M^6-72(1-b)^2g^2M^4.
\end{align*}
It is obvious that event horizons exists only for the range $0\leq b<1$.  The size of event horizon of Bardeen black hole increases with CoS parameter $(b)$ and decreases with magnetic charge parameter $(g)$. For a given value of the CoS parameter, there exists a critical value of magnetic charge beyond which the black hole horizon does not exist, as shown in Fig.(1). Event horizon correspond to $b=0$ matched with reported value of Bardeen black hole \cite{Bakhtiyor:2021}.
    \begin{figure*}[ht]
    \begin{center}
  \includegraphics[width=0.7\linewidth, height=0.45\textheight]{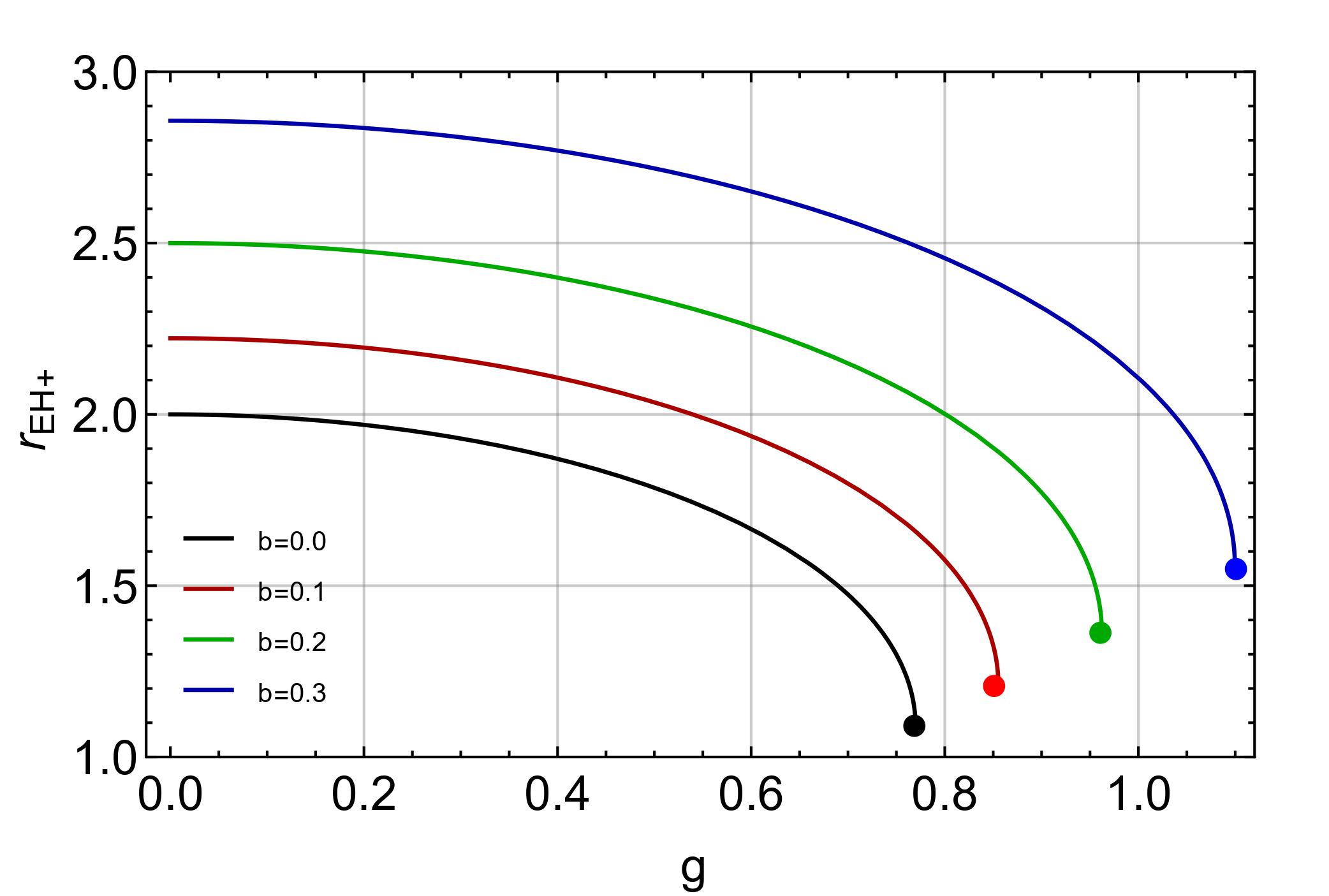}
\caption{The plot of event horizons versus magnetic charge $(g)$ for fixed value of CoS parameter, $(b)$. Circle points represent critical value of outer event horizons for fixed value of $b$.}
\label{fig:1}
\end{center}
\end{figure*}
 
\subsection{Effective potential and impact parameter}
As the null geodesic passes  near a black hole, it deviates from its original path due to strong gravitational field. In strong deflection limit, we consider light rays propagating close to the spherical photon sphere in the equatorial plane~${\theta=\frac{\pi}{2}}$. The effective potential is given by,
\begin{eqnarray}
\frac{dr}{d\tau}+W_{eff}(r)=1,
 \end{eqnarray} 
where ${\tau}$ is affine parameter, and ${W_{eff}(r)}$ is the effective potential for the Bardeen black hole in cloud of strings,
\begin{eqnarray}
W_{eff}(r)=\left[1-b-\frac{2Mr^2}{(r^2+g^2)^\frac{3}{2}}\right]\frac{u^2}{r^2};
 \end{eqnarray}
 where the ratio of angular momentum to energy ${u}=L/E$ is identified with the impact parameter.

Unstable spherical photon orbit radius is identified with the distance where deflection angle diverges and is obtained by the conditions,
\begin{align*}
 \frac{dr}{d\tau}=0, ~~~ &\frac{d^2r}{d\tau^2}=0,~~~
\frac{d^2W(r)}{d\tau^2}<0&.
\end{align*}  

Fig.(2) is an illustration of the bending of light in which a black gray circle represents a Bardeen black hole immersed in CoS. Thick horizontal lines PQ, KT and AB define the plane of source (Star), gravitational lens in sky and an observer on earth. $D_{LS}$,$D_{OL}$ and$D_{OS}$ are distances in between Lens-Source, Observer-Lens and Observer-Source respectively.  Actual light rays are incoming red rays and dotted lines are backward traces for images of the source. ${\xi}$ is an angle between the lens and the source S.
   \begin{figure}[!ht]
   \begin{centering}
  \includegraphics[width=0.40\textwidth, height=0.6\textwidth]{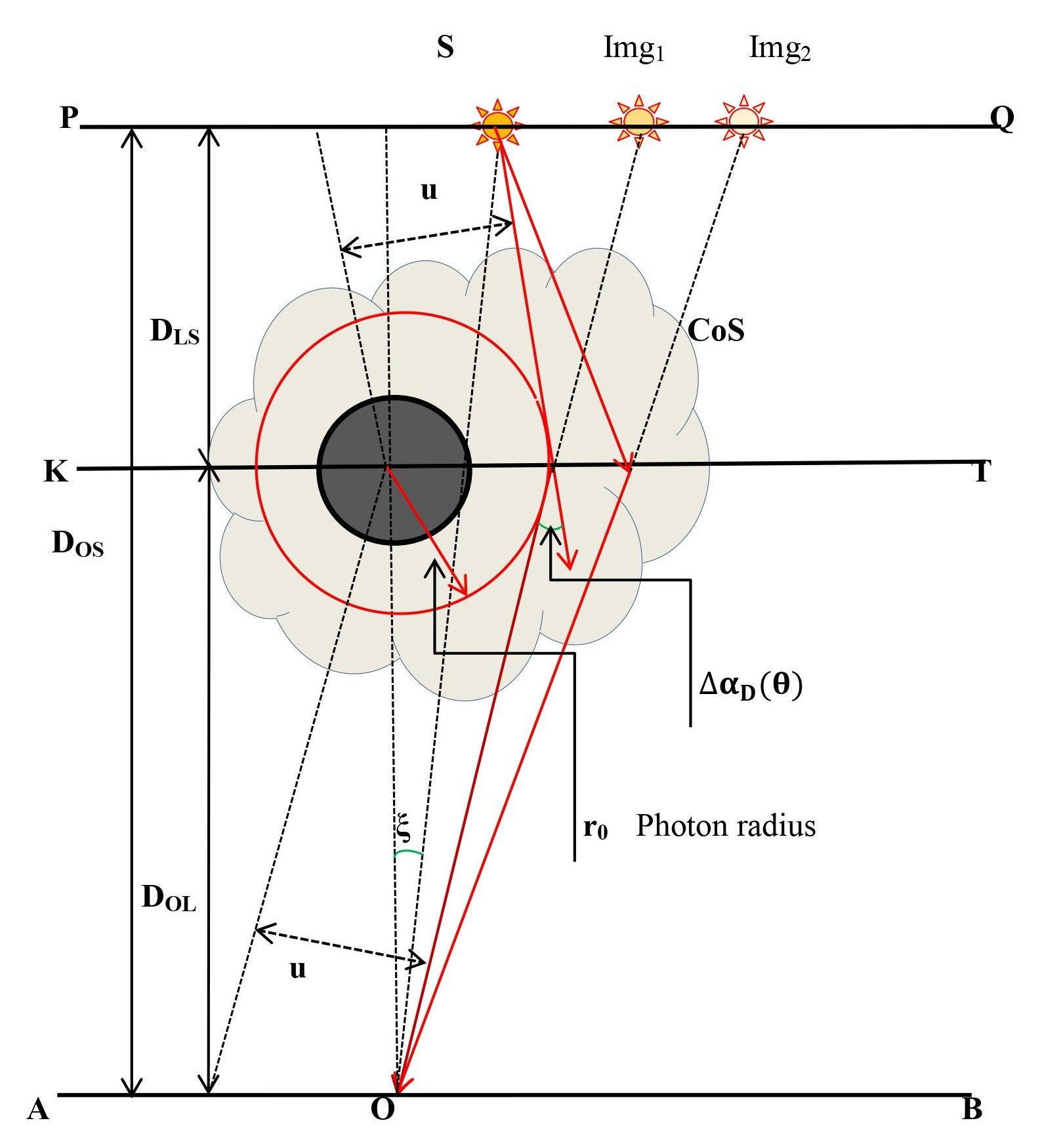}
  \qquad
    \begin{tabular}[b]{|c|c|c| c| c|}\hline 
 $r_{m}$ & b=0.0  & b=0.1  & b=0.2  & b=0.3\\ 
\hline
 g=0.0 & 3.00000 & 3.33333 & 3.75000 &4.28571 \\ 
 \hline
 g=0.1 & 2.99164 & 3.32581 & 3.74332 & 4.27987\\  
 \hline
 g=0.2 & 2.96617 & 3.30297 & 3.72308 &4.26221\\  
 \hline
g=0.3 & 2.9224 & 3.26396 & 3.6887 &4.23235\\  
 \hline
 g=0.4 & 2.85798 & 3.20715 & 3.63908 &4.18959\\  
 \hline
 g=0.5 & 2.76871 & 3.1298 & 3.57247 &4.13282\\  
 \hline
 g=0.6 & 2.64674 & 3.02721 & 3.48607 &4.06042\\  
 \hline
 g=0.7 & 2.47795 & 2.89075 & 3.37528 &3.96991\\
 \hline
 g=0.8 & 2.19830 & 2.70157 & 3.23197 &3.85746\\  
 \hline
 g=0.9 & ... & 2.39726 & 3.03927 &3.71668\\  
 \hline
 g=1.0 &... & ... & 2.74788 &3.53565\\  
 \hline
 g=1.10 &... & ... & ... &3.08654\\
 \hline
\end{tabular}
\caption{Schematic diagram of the black hole lens in cloud of strings and table of $r_{m}$ radii of unstable spherical photon orbits.}
 \end{centering}
\end{figure}
%%%%....................................
Incoming null geodesics are deflected towards black hole at fixed value of the impact parameter and makes unstable photon sphere surrounding the event horizon denoted by a black circle. The numerical values of radii of photon sphere are shown in table. Its radius increases with CoS parameter $b$ and decreases with the increase of the magnetic charge parameter $g$.\\

We  plot the effective potential and impact parameter for different values of CoS parameter in Fig.(3) for fixed values of black hole  parameters (black hole mass is taken as $M=1$ unless specified).
It is clearly observed that there exists a critical point where unstable circular orbits are possible. This defines the distance of minimum turning and the corresponding value of impact parameter is denoted by $u_{m}$. The value of impact parameter increases with the increase of the CoS parameter. Furthermore, the effective potential reaches its maximum at the largest value of angular momentum for incoming photons, indicating that this corresponds to the threshold for the formation of these unstable orbits. The graphical representation of the impact parameter as a function of the magnetic charge parameter for different values of the CoS parameter clearly shows that the impact parameter exhibits a slight decrease as the value of $g$ increases. Notably, the impact parameter reaches its maximum for higher values of the CoS parameter, emphasizing the significant role that the CoS parameter plays in influencing the system's behavior (see Fig.(3) right).
\begin{figure*}[ht]
\begin{centering}
\includegraphics[width=0.48\linewidth, height=0.35\textheight]{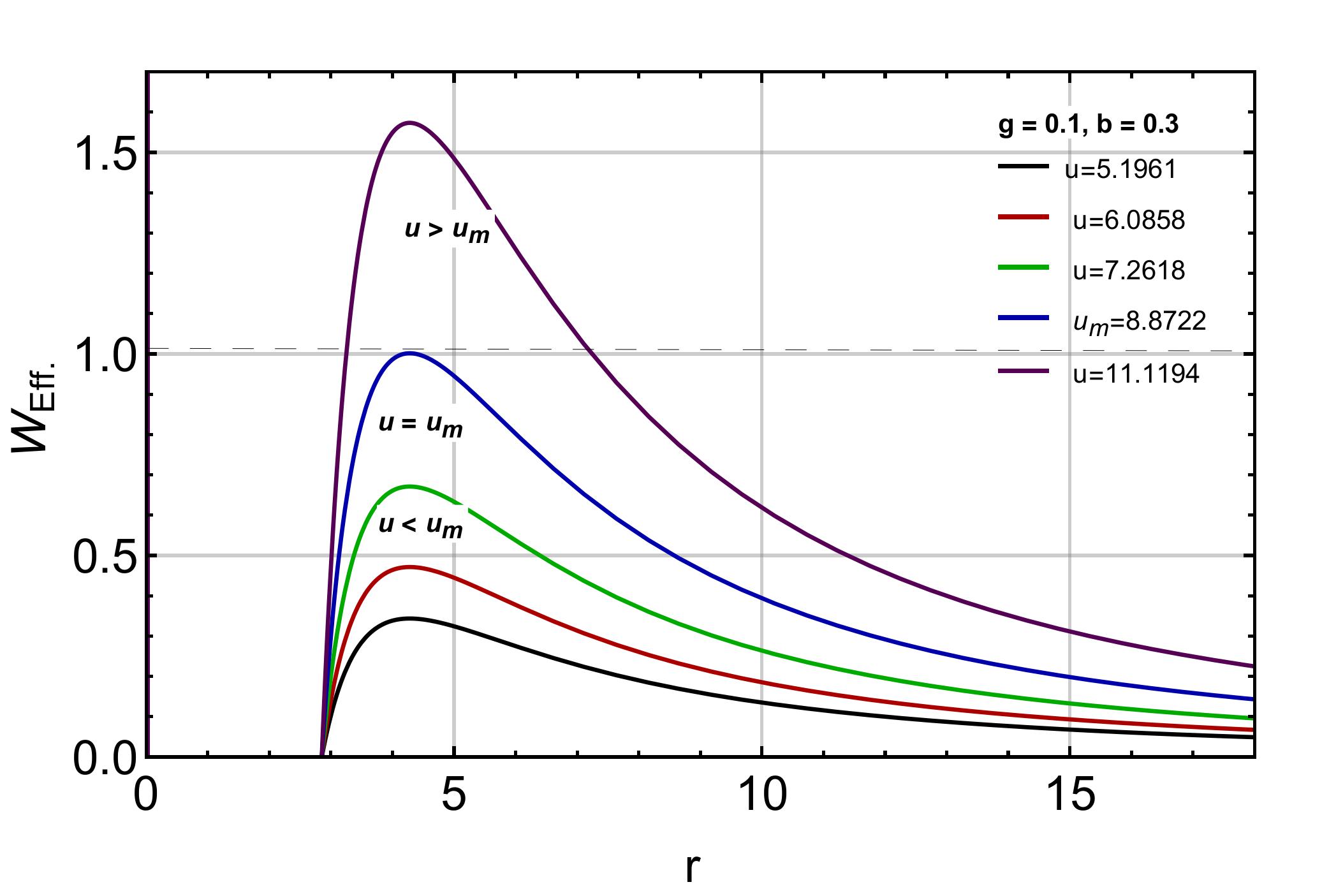}
\includegraphics[width=0.48\linewidth, height=0.35\textheight]{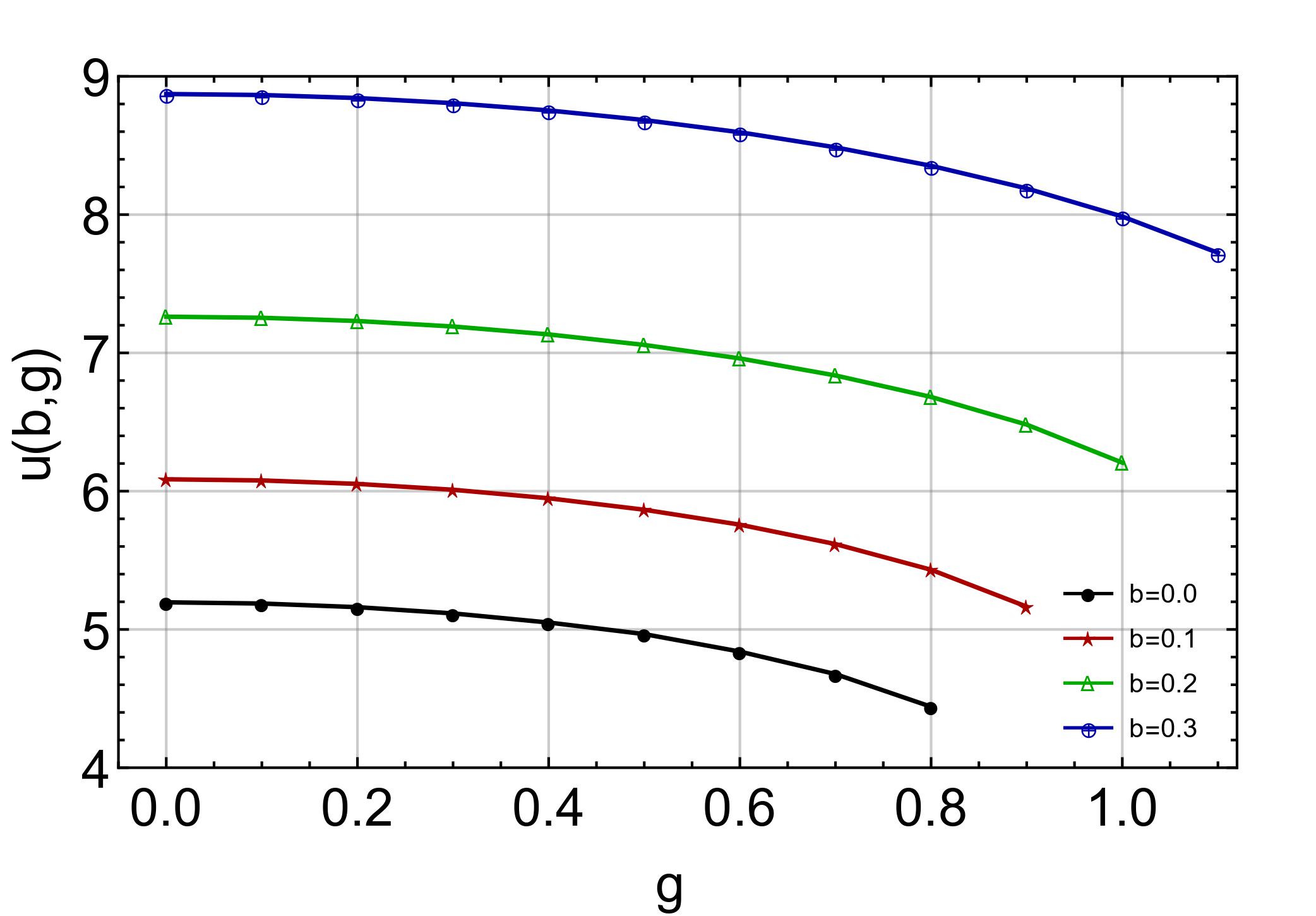}
\caption{Left plot of effective potential for fixed $g=0.1,b=0.3$ for different value $u$ and right plot of impact parameter {$u(b,g)$ }versus $g$.}
\label{fig:3}
\end{centering}
\end{figure*}

\subsection{Strong deflection limit coefficients}
 The angular shift of the photon from a black hole lens is described by the geodesic equation \cite{Chitre:1998},
\begin{eqnarray}
\frac{d\phi}{dr}=\frac{\sqrt{B}}{\sqrt{C}\sqrt{\frac{C}{C_{0}}(\frac{A_{0}}{A})-1}};
\end{eqnarray}
where subscript $0$ means that the metric functions are evaluated at some distance of minimum approach $r_{0}$ from the center of the black hole. Deflection angle is obtained by the following relation,
\begin{equation}
    \alpha_{D}(r_{0})=I(r_{0})-\pi;
\end{equation}
where,
\begin{equation}
    I(r_{0})=2\int_{r_{0}}^{\infty}\frac{d\phi}{dr}dr=\int_{r_{0}}^{\infty}\frac{\sqrt{B}}{\sqrt{C}\sqrt{\frac{C}{C_{0}}(\frac{A_{0}}{A})-1}}dr.
\end{equation}

By making a change of variable $z=\frac{A-A_{0}}{1-A_{0}}$ the above integral becomes \cite{Chitre:1998},

 \begin{equation}
      I(r_{0})=\frac{2r^{2}}{r_{0}}\int_{0}^{1}R(z,r_{0})f(z,r_{0})dz,\\
 \end{equation}
      where,

     \begin{align*}
     R(z,r_{0})=\frac{2\sqrt{AB}}{C A'}(1-A_{0})\sqrt{C_{0}},\\
      f(z,r_{0})=\frac{1}{\sqrt{A_{0}-\frac{C_{0}}{C}[(1-A_{0})z]}}.
 \end{align*}

Expanding $ f(z,r_{0})$ by Taylor's expansion upto $z^2$ which yields \cite{Ernesto},
\begin{equation}
    f(z,r_{0})=\frac{1}{\sqrt{\lambda_{1}z+\lambda_{2}z^2}},
\end{equation}
where $\lambda_{1}$ and $\lambda_{2}$ are given by the following expressions,
\begin{align*}
    \lambda_{1}=\frac{1-A_{0}}{C_{0}A_{0}'}(C_{0}'A_{0}-A_{0}'C_{0}),\\
    \lambda_{2}=\frac{(1-A_{0})^{2}}{C_{0}^{2}A_{0}^{3}}[2C_{0}C_{0}'A_{0}'^{2}-A_{0}A_{0}''C_{0}'+A_{0}A_{0}'(C_{0}''C_{0}-2C_{0}'^{2})].
\end{align*}

The lensing coefficients are given by,
\begin{equation}
    \rho=\frac{R(0,r_{ph})}{2\sqrt{\lambda_{2}}},
\end{equation}
where $r_{m}=r_{ph}$ is largest photon radius and 
\begin{equation}
    \sigma=C_{R}+\lambda_{1}ln2\delta,
\end{equation}
where $\delta=\frac{2\lambda_{2}}{A(r_{ph})}$ \& $C_{R}=I(r_{ph})$.

The expression for the bending angle can be obtained in terms of the strong deflection limit coefficient $\rho$ and $\sigma$,
\begin{equation}
 \alpha^*(u)=-\rho\log (\frac{u}{u_{ph}}-1)+ \sigma + \mathcal{O} (u-u_{ph})...... .
\end{equation}
 
\begin{figure*}[ht]
\begin{tabular}{c c }
\includegraphics[width=0.45\linewidth, height=0.30\textheight]{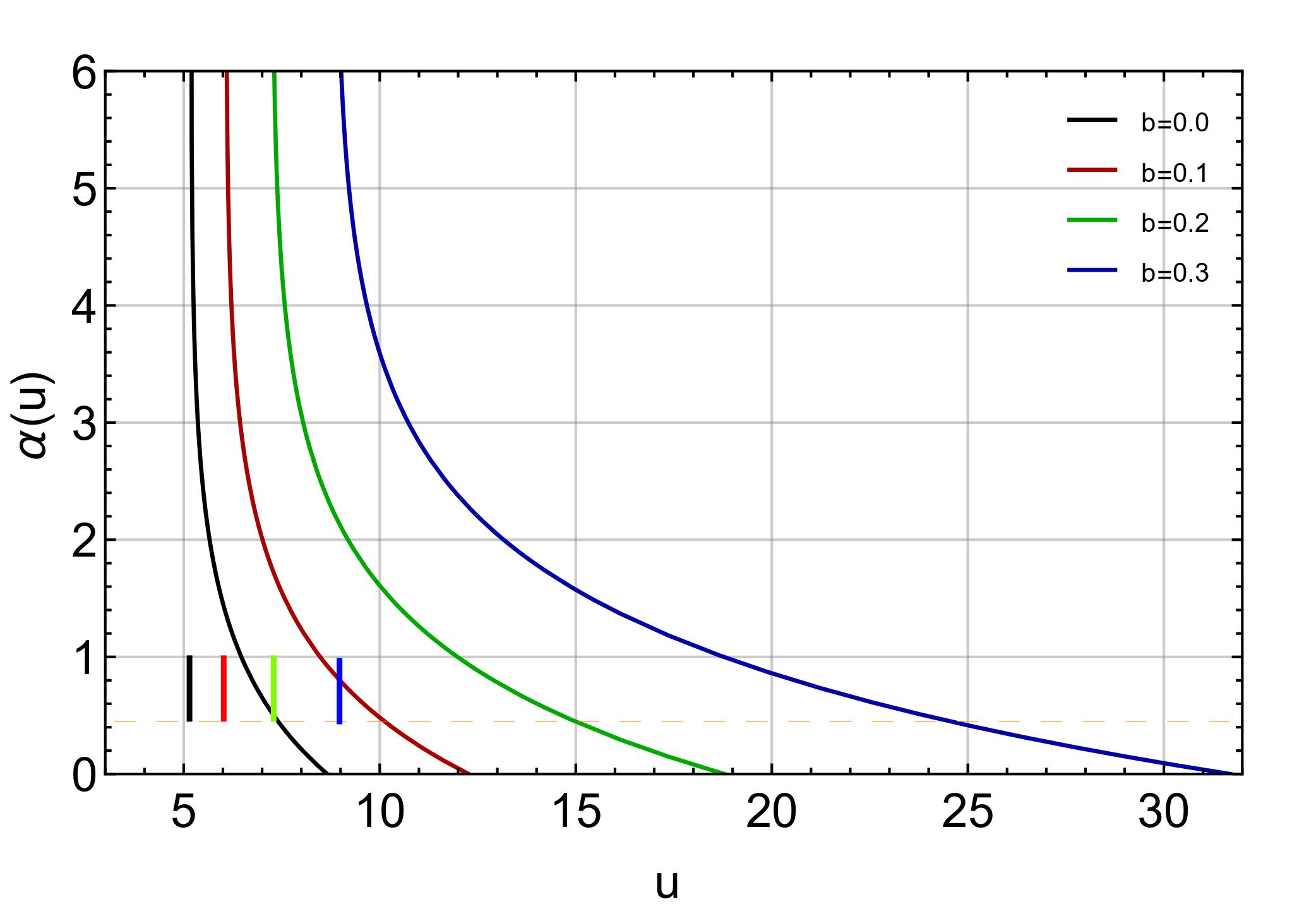}
\includegraphics[width=0.45\linewidth, height=0.30\textheight]{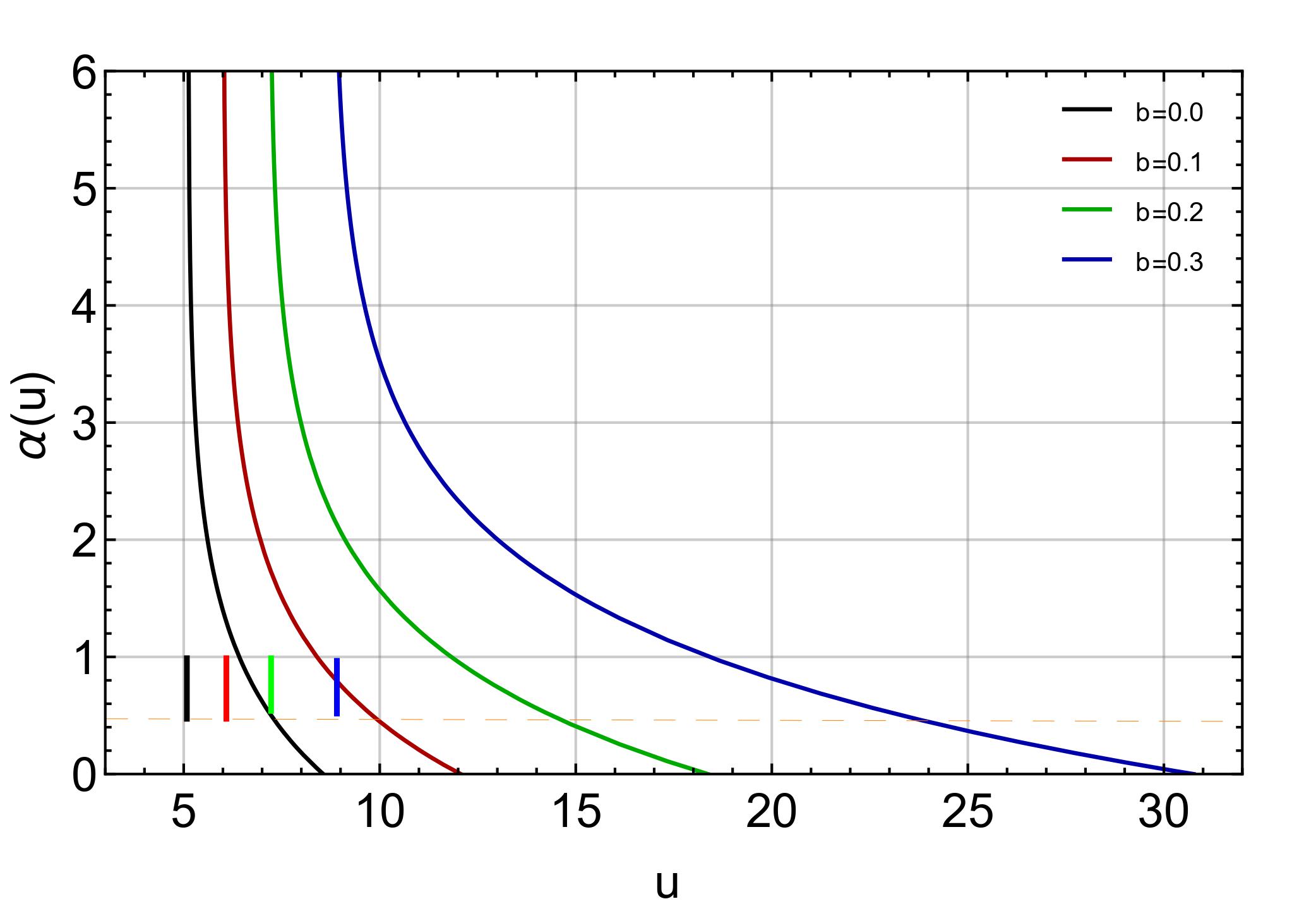}\\
\end{tabular}
\caption{The plot of deflection angle as function of impact parameter, $u$ for fixed value of magnetic charge parameter, g=0.1 (left) and g=0.3 (right)}.
\label{fig:4}
\end{figure*}

\begin{table}[ht]
\centering
\begin{tabular}{ |p{5cm}||p{5cm}||p{5cm}|  }
\hline
 ~~~~~~~CoS parameter b & ~~~~~~~ $u_{m}$ (g=0.0)  & ~~~~~~~~ $u_{m}$ (g=0.7)\\
    \hline
    ~~~~~~~~~~~~~~   0.0   & ~~~~~~~~~~~~~~ 5.081   & ~~~~~~~~~~~~~~  4.677  \\
    \hline
   ~~~~~~~~~~~~~~   0.1 & ~~~~~~~~~~~~~~ 6.088   & ~~~~~~~~~~~~~~  5.617  \\
    \hline
   ~~~~~~~~~~~~~~   0.2 & ~~~~~~~~~~~~~~ 7.296   & ~~~~~~~~~~~~~~  6.836   \\
    \hline
  ~~~~~~~~~~~~~~  0.3 & ~~~~~~~~~~~~~~ 8.975  & ~~~~~~~~~~~~~~ 8.486   \\ 
    \hline
\end{tabular}
\caption{ Critical values of the Impact parameter $u_{m}$ for different values of $b$ and $g$.}
\end{table}

The variation in strong deflection angle are shown in Fig.(4). Plots below have been drawn in comparison with Bardeen black hole $b=0$. Latelier parameter (CoS) yields large divergence of deflection angle corresponding to a given value of impact parameter $u$. It is clear that effect of magnetic charge parameter is less significant than CoS parameter. The critical impact parameter for different values of $b$ and $g$ are listed in the table (1). 

\section{Magnifications, Relativistic Einstein's Rings and EHT constraints}
In this section, we investigate the image magnification and relativistic Einstein rings of the black hole\footnote{The term relativistic Einstein rings was coined by Virbhadra and Ellis \cite{KS:2000}}.  Let $\xi$ and $\theta$ measures the angular position of the  source  (S) and the image (I) respectively (see Fig.2). These are related by the well-known Virbhadra-Ellis lens equation equation \cite{KS:2000,KS:2009},

\begin{eqnarray}
    \tan{\xi_{S}}=\tan{\theta_{I}}-\frac{D_{SL}}{D_{SO}}[\tan{\theta_{I}}+\tan(\alpha-\theta_{I})].
\end{eqnarray}

Here, $D_{LS}$, $D_{SO}$ and $D_{OL}$ are the distances between source-lens, source-observer and observer-lens.

If we consider perfect alignment, ${\tan{\theta}}$=${\theta}$ then the above equation reduces to,
\begin{eqnarray}
    \xi_{S}=\theta_{I}-\frac{D_{LS}}{D_{OS}}\Delta{\alpha_{n}}.
    \end{eqnarray}
     In lens equation the term $\Delta{\alpha_{n}}$ is the deflection angle (DA) of n-th image. 

    \begin{eqnarray}
    \theta_{n}=\theta^{0}_{n}+\frac{({D_{OL}+{D_{LS}}})}{D_{LS}}.\frac{u_{m}e_{m}}{D_{OL}\rho}(\xi_{S}-\theta^{0}_{n}).
    \end{eqnarray}
    Here, ${\theta}^{0}_{n}$ is some reference 
    initial position of the image and true  for both sides of lens. Using, $\xi_{S}=0$ and $D_{OS}=2D_{OL}$, we obtain 
    the position of the Einstein ring given by,  
     \begin{eqnarray}
    \theta^{E}_{n}=\frac{u_{m}}{D_{OL}}(1+e_{n})
     \end{eqnarray}
    \text{ where $e_n$} is the 
    \begin{align*}
    & e_{n}=\exp{\frac{\sigma}{\rho}-\frac{2n\pi}{\rho}}.
      \end{align*}
      It is clear that Einstein's ring size decrease for higher order of $n$, where $n$ is an integer except $0$. Also it decrease as the distance between lens and the observer increase and becomes larger in case of massive black hole. In the following sections we will discuss magnification and analyzed outermost  Einstein ring in the equatorial plane of the black hole lens.

\subsection{Magnification}

Magnification of relativistic images is an important observable and it is defined as the ratio of two solid angles made by the image and the source on the observer's eye, respectively \cite{Blandford:1992}. It is determined by the following relation,
    \begin{eqnarray}
       \mu=\left(\frac{\sin\xi_S\partial\xi_S}{\sin\xi_\theta\partial\xi_\theta}\right)^{-1}.
       \end{eqnarray}
       ${\xi_{S}}$ and ${\xi_{\theta}}$ are the angular size of the source and the image, respectively.

        The expression for the magnification in terms of SDL coefficients and the position of the relativistic image and the source is given by, \cite{Schneider:1992,Mancini:2004,Parlik:2004,Islam:2022},
        \begin{eqnarray}
         \mu_{n}=\frac{1}{\xi\rho}\frac{{u_{ph}}^2}{{D_{OL}}^{2}}{e_{n}(1+e_{n}})(1+\frac{D_{OL}}{D_{LS}}),
         \label{am}
\end{eqnarray}
\vspace{1cm}
where ${\xi}$ and ${\rho}$ represents the source position and first lensing coefficient. 

In Fig.(5) we plot absolute magnification (A.M.)  with respect to source position $\xi$ for $SgrA^{*}$ for different values of the black hole parameters $b$ and $g$. Magnification of first primary image is denoted by ${\mu_{P1Img}}$.
%\newpage

\begin{figure*}[ht]
\begin{centering}
\begin{tabular}{c c}
\includegraphics[width=0.45\linewidth, height=0.25\textheight]{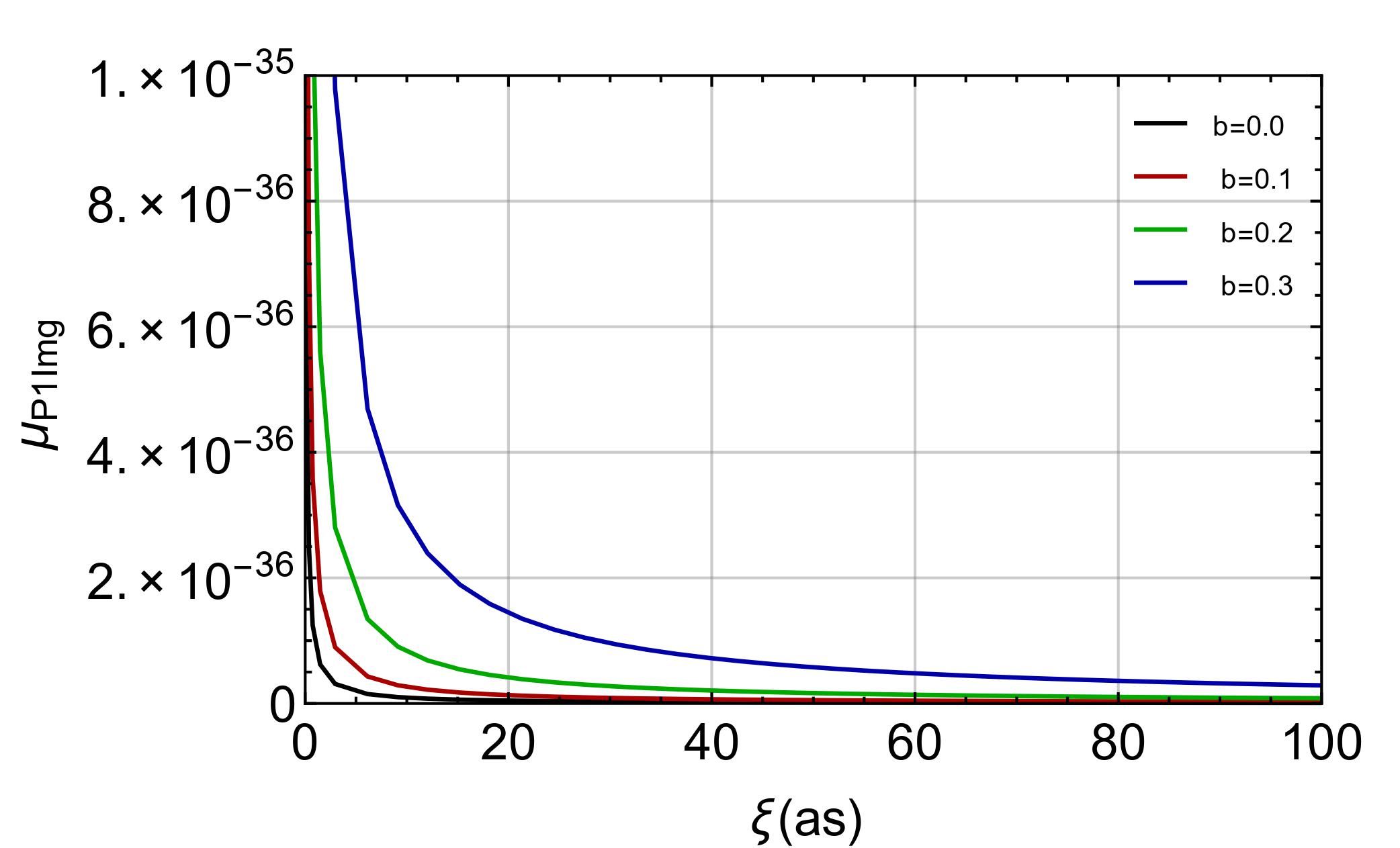}
\includegraphics[width=0.45\linewidth, height=0.25\textheight]{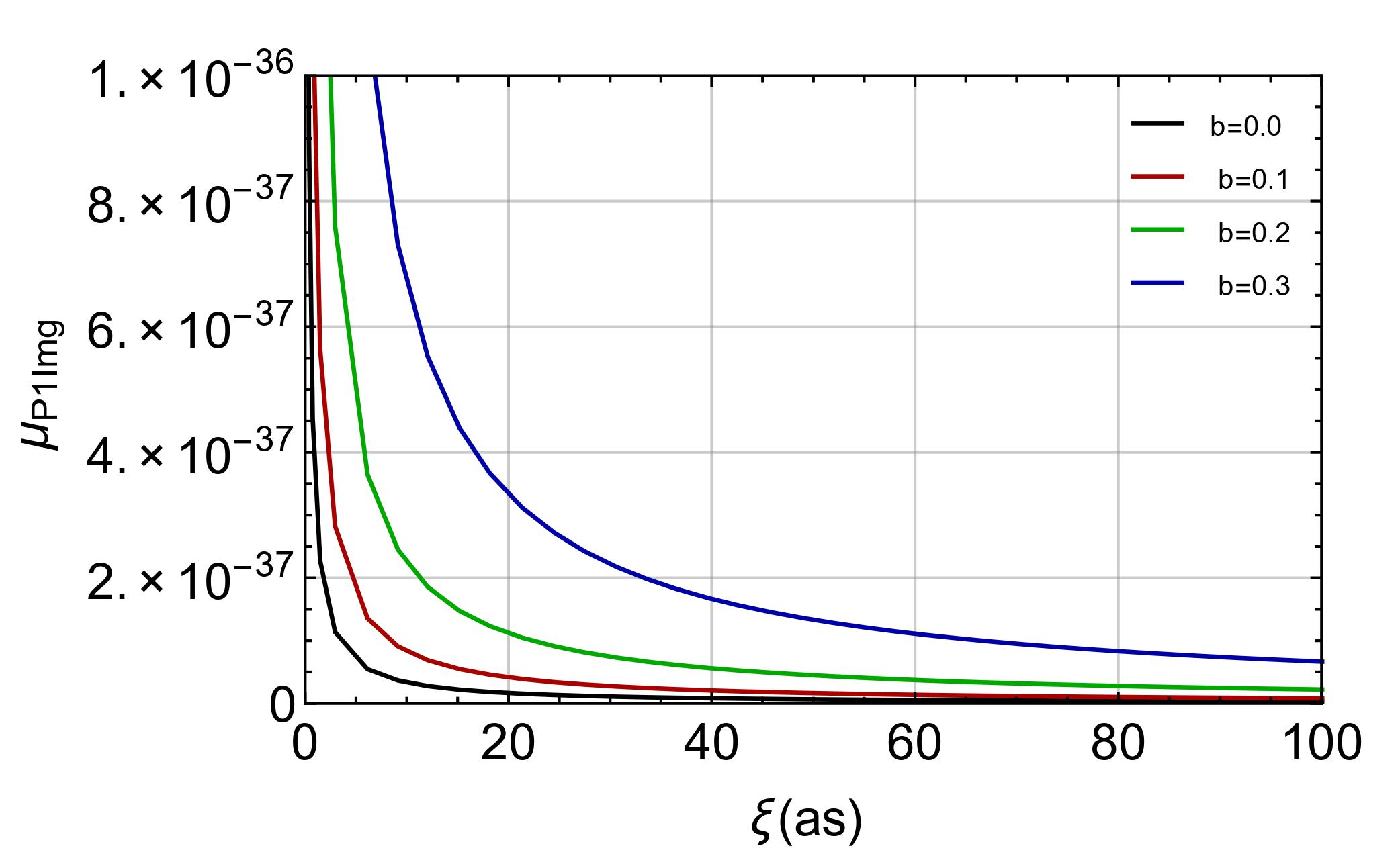}\\
\end{tabular}
\caption{A.M. of first primary image for fixed value of $g=0.1$ (left) and $g=0.3$ (right), with respect to position of the source (in arcsec) for $SgrA^{*}$.}
\end{centering}
\label{fig:5}
\end{figure*}

\begin{figure*}[ht]
\begin{centering}
\begin{tabular}{c c}
\includegraphics[width=0.45\linewidth, height=0.25\textheight]{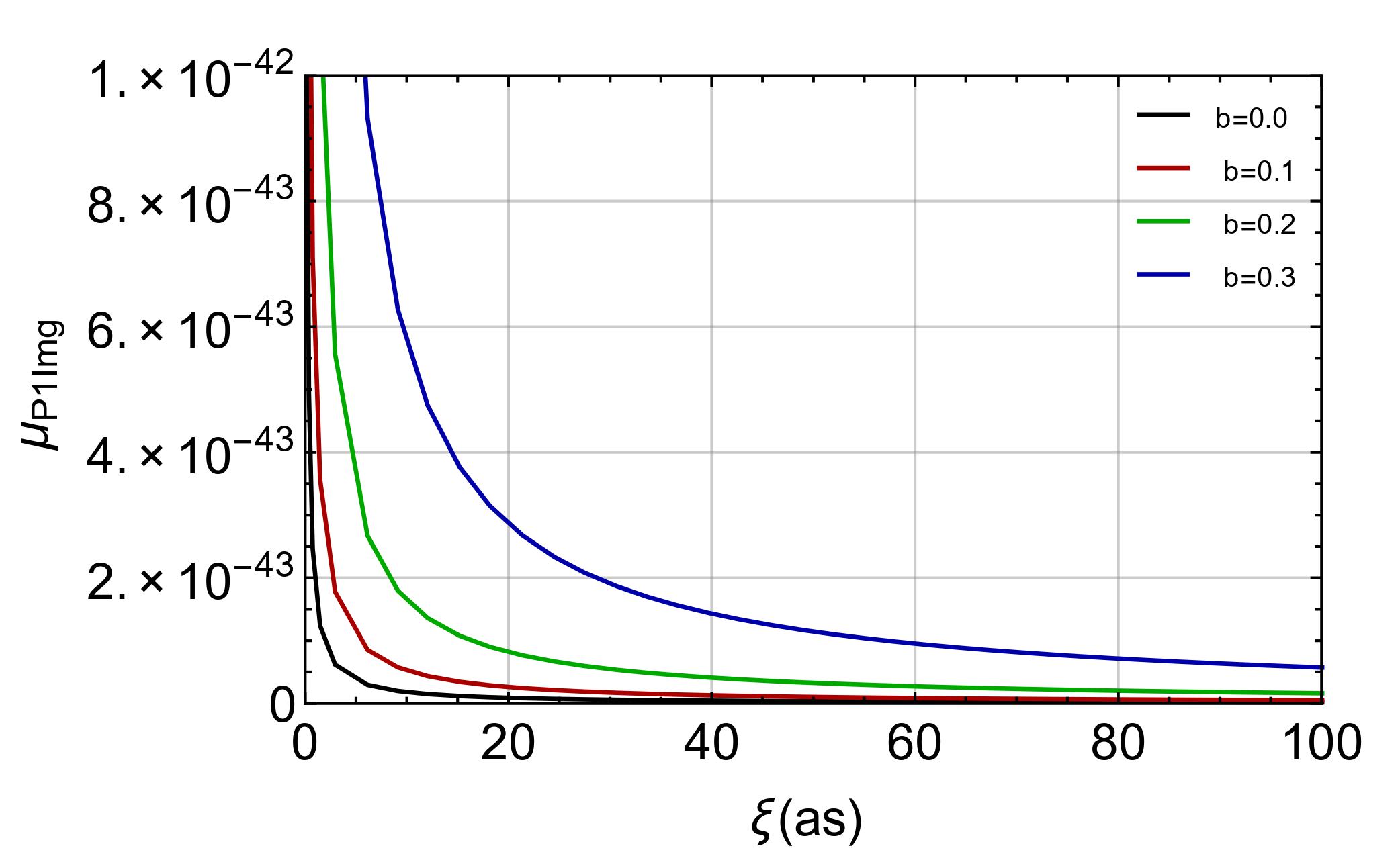}
\includegraphics[width=0.45\linewidth, height=0.25\textheight]{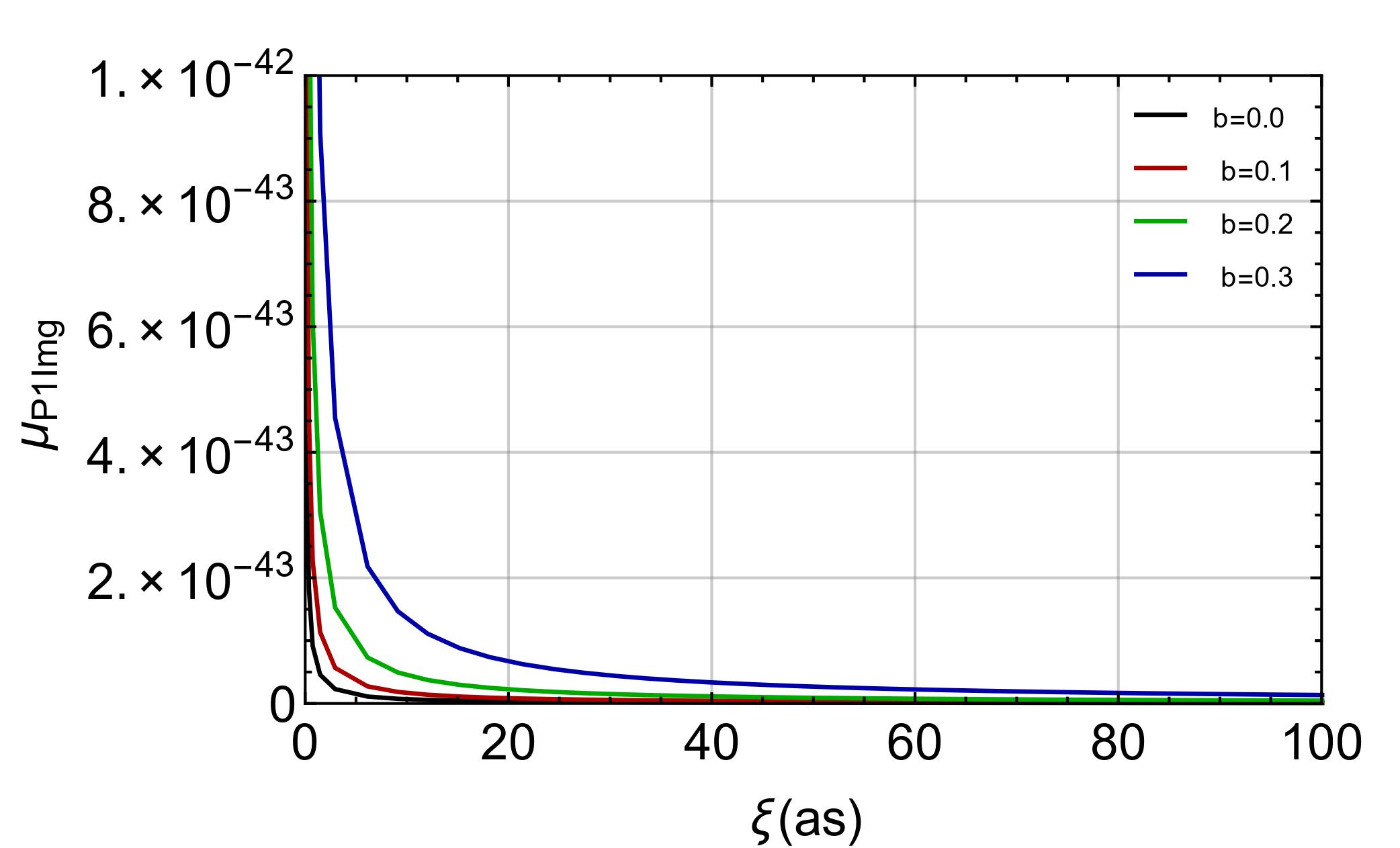}\\
 \end{tabular}
\caption{A.M. of first primary image for fixed value of $g=0.1$ (left) and $g=0.3$ (right) with respect to position of source (in arcsec) for $M87^{*}$.}
\end{centering}
\label{fig:6}
\end{figure*}
It is evident from equation (\ref{am}), ${\xi}$ approaches to zero near optic axis. At this point magnification diverges and all caustic points lined up on optic axis and one obtains magnified image. In Fig. (6) we have shown absolute magnification of supermassive black hole  $M87^{*}$ and found that its absolute magnification is small compared to $SgrA^{*}$ in Fig.(5). In both cases, first order images are highly magnified if we turn on the CoS parameter, $b$ and magnetic charge, $g$. Hence, CoS acts as magnifier, while magnetic charge parameter acts as demagnifier for first order relativistic images. It is clear that absolute magnification decreases with respect to source position. Source and observer are in along optic axis so relativistic images occur symmetrically on both sides of lens at minimum impact angle.

%\newpage

\subsection{Relativistic Einstein's Rings}
The relativistic Einstein ring is formed when the gravitational field of a massive foreground object, such as a galaxy or a black hole, perfectly aligns with a distant light source and the observer \cite{KS:2000}. In this precise alignment, the gravitational lens effect causes light from the background source to bend around the foreground object, forming a symmetrical ring of light. This phenomenon is a direct consequence of General Relativity, where the massive object's gravitational field warps the fabric of spacetime, causing the light to follow a curved path.
\begin{table}[!ht]
  \begin{centering}
  \begin{tabular}{|c|c|c|c|c|c|c|c|c|c|}
    \hline
    \hline
    \hline
      \diagbox[width=0.7 \textwidth/10+4\tabcolsep\relax, height=1.5cm]
      { $g$ }{$CoS$} & \multicolumn{3}{c|}{$b = 0.0$} & \multicolumn{3}{c|}{$b = 0.1$} & \multicolumn{3}{c|}{$b = 0.3$} \\ \hline
    g & $\theta_{\infty}$ & $\theta_{ring}$ & $\theta_{sh}$
    & $\theta_{\infty}$ & $\theta_{ring}$ & $\theta_{sh}$
    & $\theta_{\infty}$ & $\theta_{ring}$ & $\theta_{sh}$ \\ 
    \hline
 0.0 & 24.9415 & 24.9727 & 49.8830  & 29.2118 
 & 29.2118 & 58.4236  & 42.5868 & 43.4823 & 86.3250\\
 \hline
 0.1 & 24.8998 & 24.9315 & 49.7996  & 29.1723 & 29.2810 & 58.3446  
  & 42.5520& 43.1275 & 86.2546\\
 \hline
 0.3 & 24.5565 & 24.5930 & 49.1130 & 28.8486
 & 28.9336 & 57.6972 & 42.4469 & 43.0358 & 86.0716\\
 \hline
 0.5 & 23.8391 & 23.8919 & 47.6782 & 28.1558
 & 28.2633 & 56.3116 & 41.6822 & 42.2605 & 84.5210\\
 \hline
 0.7 & 22.4505 & 22.5881 & 44.9010 & 26.9636 
 & 27.1507 & 53.9272 &  40.7356 & 41.3000 & 82.6000\\
 \hline
  \end{tabular}
\caption{This table shows angular measures of $SgrA^{*}$ $(in ~~ {\mu}arcsec)$.}
\end{centering}
\end{table}
The radius of the Einstein ring is proportional to the mass of the foreground object and inversely proportional to its distance from the observer. There are infinitely many apparent images of actual source due to strong lensing. These apparent images of a single source forms rings and the outermost ring is called Einstein ring (n=1). The successive inner ring are called relativistic Einstein rings. We determine the position of nth relativistic images and the Einstein ring.
%\newpage
\begin{table}[!ht]
  \begin{centering}
  \begin{tabular}{|c|c|c|c|c|c|c|c|c|c|}
    \hline
    \hline
    \hline
      \diagbox[width=0.7 \textwidth/10+4\tabcolsep\relax, height=1.5cm]
      { $g$ }{$CoS$} & \multicolumn{3}{c|}{$b = 0.0$} & \multicolumn{3}{c|}{$b = 0.1$} & \multicolumn{3}{c|}{$b = 0.3$} \\ \hline
    g & $\theta_{\infty}$ & $\theta_{ring}$ & $\theta_{sh}$
    & $\theta_{\infty}$ & $\theta_{ring}$ & $\theta_{sh}$
    & $\theta_{\infty}$ & $\theta_{ring}$ & $\theta_{sh}$ \\ 
    \hline
0.0 & 20.7846 & 20.8106 & 41.5692 & 24.3432 & 24.4072 & 48.6864 & 35.4890 & 35.9689 & 71.9378\\
 \hline
 0.1 & 20.7763 & 20.7763 & 41.5526 & 24.3749 & 24.3749 & 48.7498 & 35.4600 & 35.9396 & 71.8792\\
 \hline
 0.3 & 20.4637 & 20.4941 & 40.9275 & 24.0405 & 24.1113 & 48.0810 & 35.2246 & 35.7133 & 71.4266\\
  \hline
 0.5 & 19.8659 & 19.9099 & 39.7318 & 23.4635 & 23.5529 & 46.9270  & 34.7352 & 35.2171 & 70.4243\\
  \hline
 0.7 & 18.7087 & 18.8234 & 37.4175 & 22.4696 & 22.6255 & 44.9392 &  33.9463 & 34.4172 & 68.8344\\
 \hline
  \end{tabular}
\caption{This table shows angular measures of $M87^{*}$ $(in ~~ {\mu}arcsec)$.}
\end{centering}
\end{table}
If the lens is very far away from the observer then, we calculate the angular radius of the Einstein ring as given in tables (2) and (3) corresponding  to different CoS and magnetic charge parameters for $SgrA^{*}$ and $M87^{*}$ black holes respectively.
\begin{figure*}[!htb]
\begin{center}
\begin{tabular}{c c }
\includegraphics[width=0.5\linewidth]{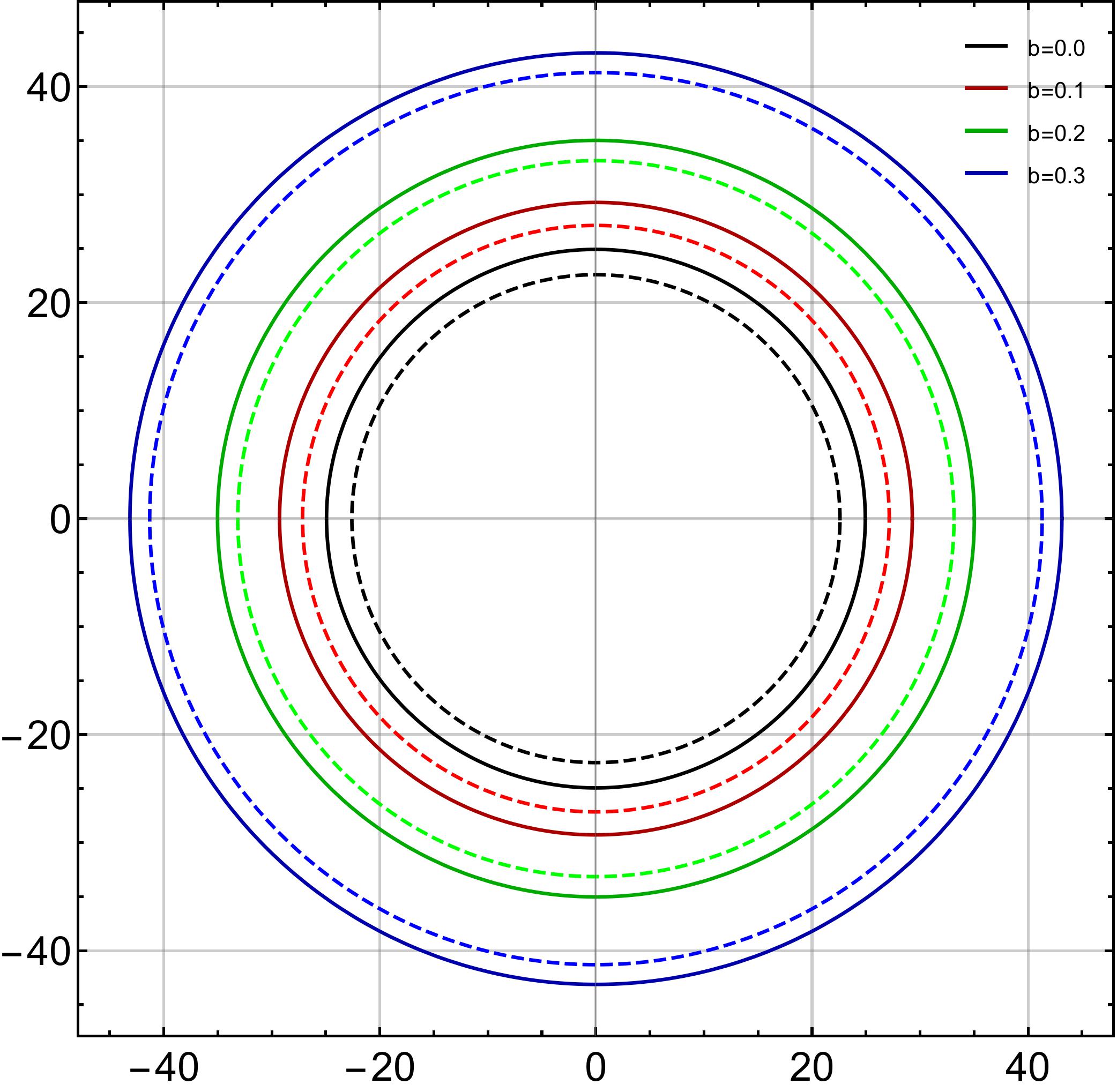}
\end{tabular}
\caption{The plot of  Einstein ring for fixed value of magnetic charge parameter, g=0.1 (solid circle) and g=0.7 (dotted circle)  for $SgrA^{*}$ in observer's sky.}
\label{fig:7}
\end{center}
\end{figure*}
%\newpage
\begin{figure*}[!ht]
\begin{center}
\includegraphics[width=0.5\linewidth]{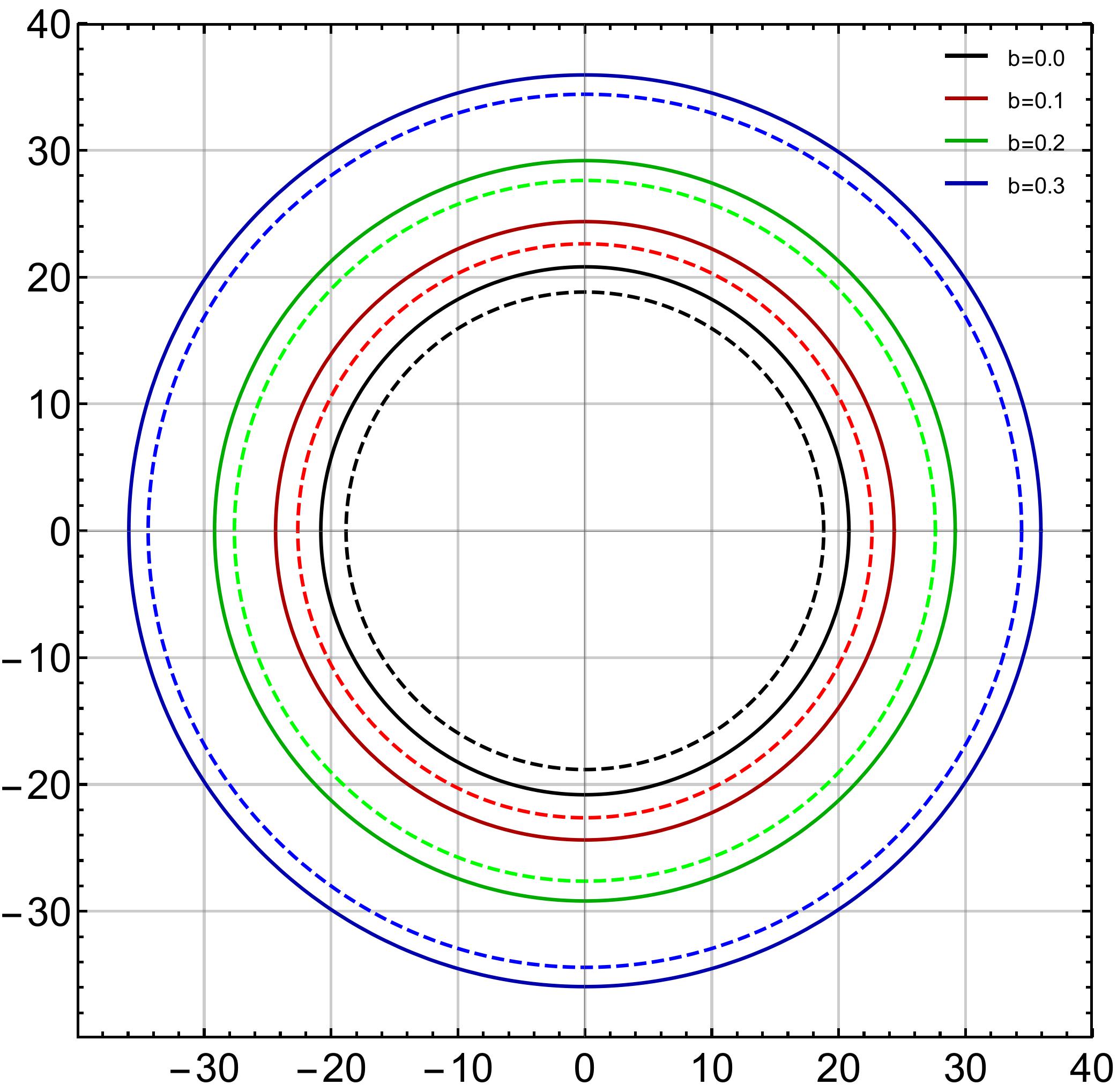}
\caption{The plot of  Einstein ring for fixed value of magnetic charge parameter, g=0.1 (solid circle) and g=0.7 (dotted circle) for $M87^{*}$ in observer's sky.}
\label{fig:8}
\end{center}
\end{figure*}
It is clear that the angular radius of the Einstein ring increases in the presence of cloud of strings and decreases due to magnetic charge. In Figs. (7) and (8), we plot the Einstein ring in two-dimensional parameter space of the observer's sky for two astrophysical black hole $Sgr A^{*}$,and $M87^{*}$ respectively. The Einstein rings are well separated. Thus, the Einstein rings may be used to differentiate astrophysical black holes present in CoS spacetime. In the next section, we shall estimate the shadow size using these rings and compare with previously reported shadows of astrophysical black hole.\\

\subsection{EHT constraints of shadow of Bardeen Black hole in cloud of strings (CoS) in the strong field limit}
In this section, we constrain CoS parameter with EHT observation of two astrophysical black holes $ SgrA^{*}$ and $M87^{*}$ \cite{Mancini:2004,
JKSG:2022}. It may appear misleading at first sight that Sgr A* and M87* represent primordial black holes surrounded by cloud of strings. However, the accretion dynamics and the interaction of the black hole with its surrounding environment may create conditions that mimic or preserve exotic matter distributions akin to string clouds \cite{Patricio:1983,Hindmarsh:1994,Davis:2005,Ganguly:2014cqa,Vachaspati:2015,Pierre:2020}.
 
 One may constrain CoS parameter of Bardeen black hole using these signals.
The shadow of the black hole is given by $\theta_{S}=2\theta_{\infty}$, where $\theta_{\infty}$ is angular position of the image in limiting case ($n$ approaches to infinity).

\begin{figure*}[ht]
\centering
\begin{tabular}{c c }
\includegraphics[width=0.48\linewidth, height=0.35\textheight]{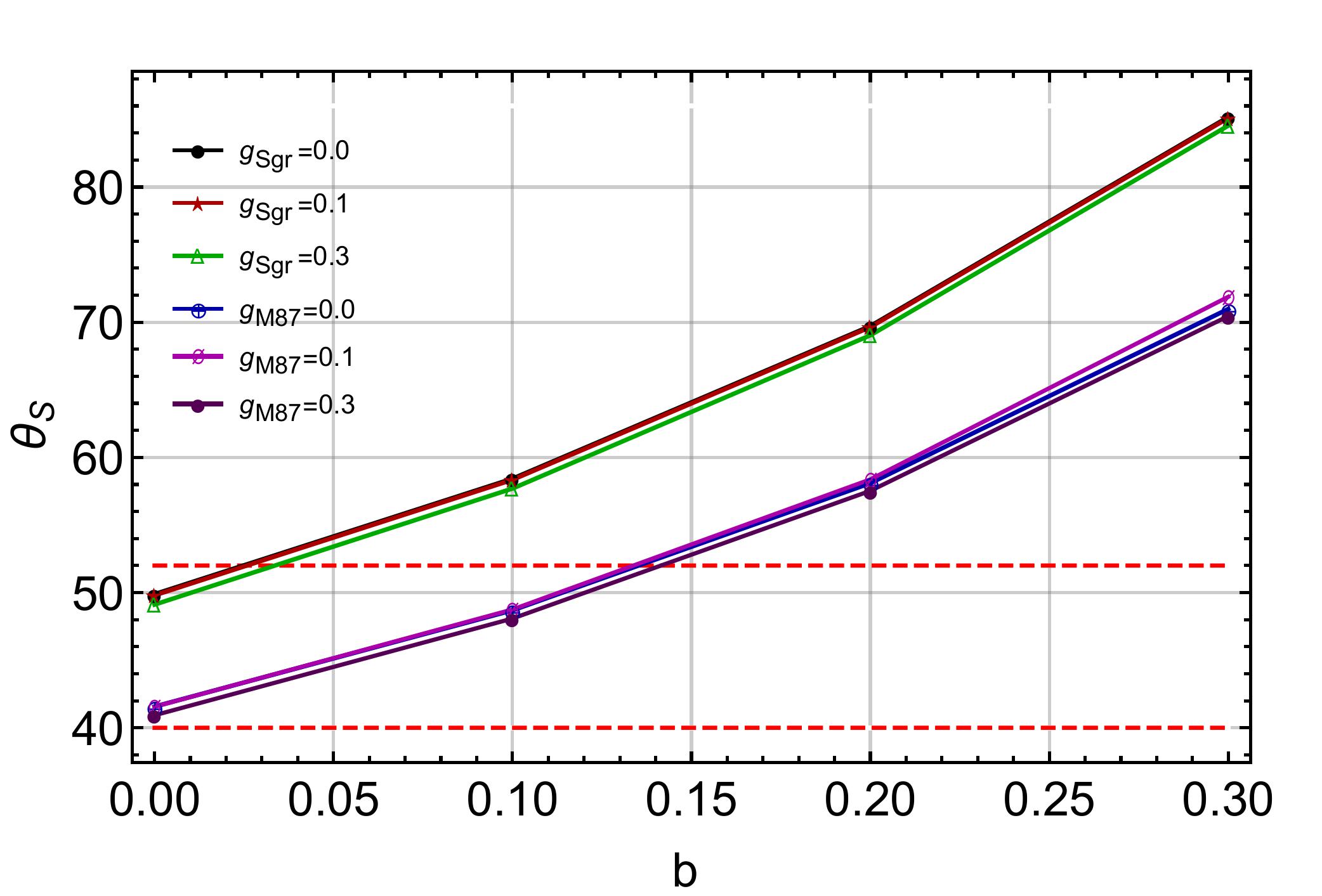}\hspace{0.5cm}
\includegraphics[width=0.48\linewidth, height=0.35\textheight]{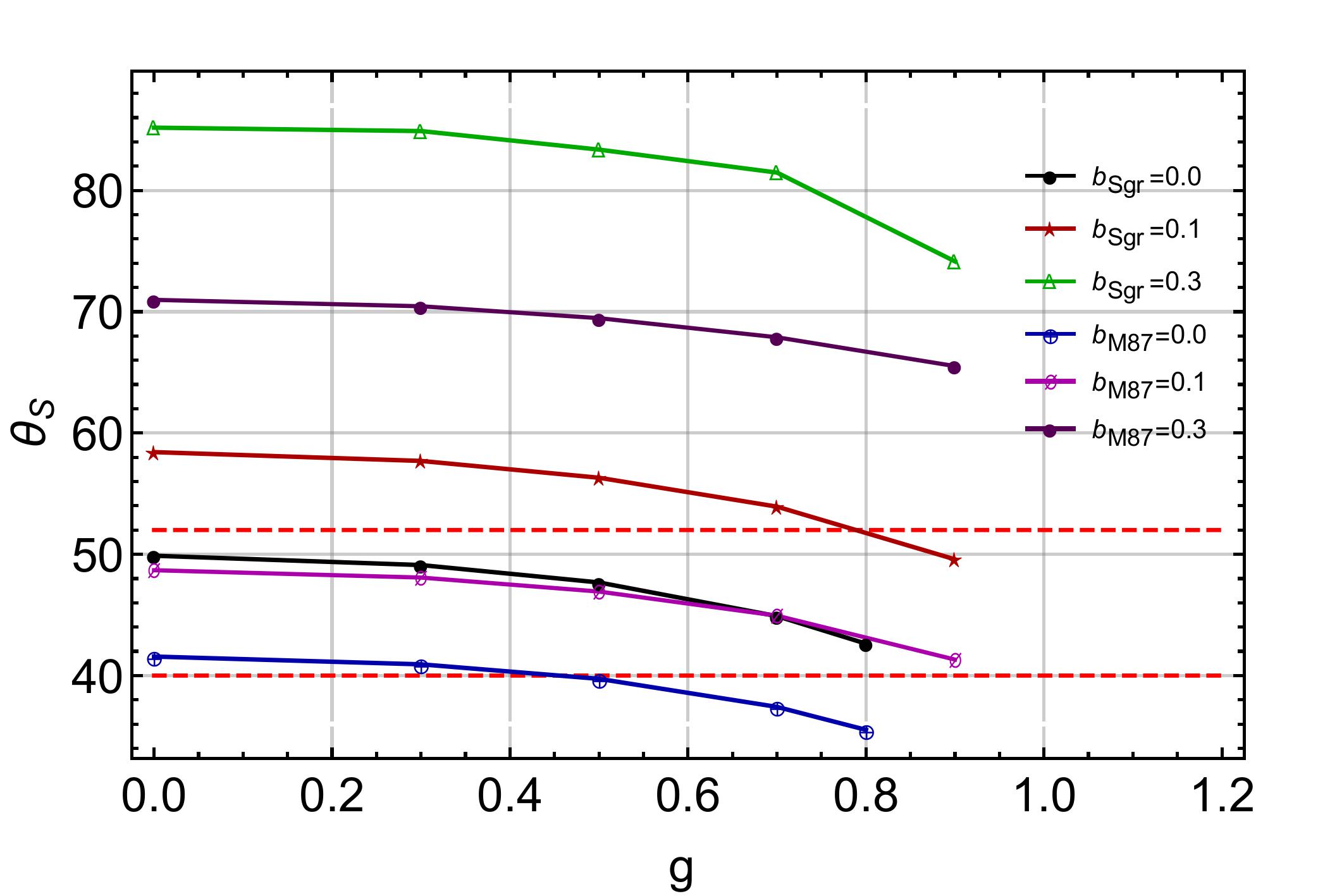}\\
\end{tabular}
\caption{The plot of constrained angular diameter of shadows versus $b$  and $g$ for fixed value of magnetic charge, $g$ (left) and CoS parameter, $b$ (right).}
\label{fig:9}
\end{figure*}
In Figs.9, we plot angular diameter (in $\mu as$) of the black hole shadow for fixed values of $g$ and $b$. Horizontal red dashed lines represent the EHT observed shadow range for astrophysical black holes $Sgr A^{*}$ and $M87^{*}$. The CoS parameter lies in range $0<b<0.09$ and $0<b<0.19$ for $Sgr A^{*}$ and $M87^{*}$ respectively. Similarly the limit for $g$ are $0.8$ and $0.9$ at constant value of $b$=$0$ and $0.1$ for $Sgr A^{*}$ and $M87^{*}$ respectively.
%%.......................................................
\newpage
\section{Conclusions} In this paper, we have analyzed regular Bardeen black hole in CoS as a gravitational lens in strong field limit. Lensing from a black hole depends on the impact parameter and we notice that the impact parameter increases with CoS parameter, $b$. The lensing produces multiple images known as relativistic rings. The magnification of the relativistic images decreases with respect to source position while radius of the ring increases with CoS parameter.  However, it is not possible to distinguish among relativistic ring, black hole shadow and a photon ring. The angular diameter of shadow decreases with magnetic charge parameter, but increases in the presence of cloud of string. This trend is identical to the rings formed by lensing. In this work we have also obtained constraints on CoS parameter for two astrophysical black holes $Sgr A^{*}$ and $M87^{*}$, in order to explore the observational signatures of cloud remnant that may still exist in our universe.

 Here, we have considered Bardeen black hole in CoS, as a gravitational lens and the work has significant implications for detection of primordial black holes and signature of exotic CoS in the very early Universe \cite{Víctor:2021,Arbey:2024}. It is plausible to extends this work to include the effects of  dark matter \cite{Ashima:2024} and rotating Bardeen black holes in CoS \cite{SS:2024,Tian:2023,Yang:2024}. The investigations can also be extended for black holes in  modified theories of gravity like f(R), f(Q) and f(R,T) \cite{Vishwakarma:2024}. Recently, gravitational wave background of primordial origin related to comic strings \cite{Domènech} is reported  and it would be interesting to relate the primordial gravitational wave background with the existence black holes in CoS.  
 
 %%%%%%%%%%%%%%%%%%%%%%%%%%%%%%%%%%%%%%%%%%%%%%%%% 

 \section*{Acknowledgment(s)}  
  BKV acknowledges the financial support from UGC fellowship.  

%%%%%%%%%%%%%%%%%%%%%%%%%%% 

\end{document}